
\input harvmac

\def\sdtimes{\mathbin{\hbox{\hskip2pt\vrule height 4.1pt depth -.3pt width
.25pt \hskip-2pt$\times$}}}
\def\nextline{\unskip\nobreak\hskip\parfillskip\break}
\def\sst{\scriptscriptstyle}
\def\frac#1#2{{#1\over#2}}
\def\coeff#1#2{{\textstyle{#1\over #2}}}
\def\hf{\coeff12}

\def\vev#1{\left\langle #1 \right\rangle}
\def\journal#1&#2(#3){\unskip, \sl#1~\bf #2\rm(19#3)}
\def\andjournal#1&#2(#3){\sl#1~\bf #2\rm(19#3)}
\def\ie{{\it i.e.}}
\def\eg{{\it e.g.}}
\def\bra#1{\left\langle #1\right|}
\def\ket#1{\left| #1\right\rangle}
\def\tr{{\rm tr}}
\def\det{{\rm det}}
\def\exp{{\rm exp}}
\def\log{{\rm log}}

\def\One{{1\hskip -3pt {\rm l}}}
\def\ephi{$e^{\gamma\phi}$}

\catcode`\@=11
\def\slash#1{\mathord{\mathpalette\c@ncel{#1}}}
\overfullrule=0pt
\def\steepslash{\c@ncel}
\def\frac#1#2{{#1\over #2}}

\def\DD{{\cal D}}

\def\FF{{\cal F}}

\def\HH{{\cal H}}

\def\OO{{\cal O}}

\def\ZZ{{\cal Z}}
\def\lam{\lambda}
\def\eps{\epsilon}

\def\p {\partial}
\def\t{{\bf t}}

\def\P{{\bf P}}
\def\Q{{\bf Q}}
\def\ptilde{\widetilde\P}
\def\qtilde{\widetilde\Q}
\def\inbar{\,\vrule height1.5ex width.4pt depth0pt}
\def\IB{\relax{\rm I\kern-.18em B}}
\def\IC{\relax\hbox{$\inbar\kern-.3em{\rm C}$}}
\def\IP{\relax{\rm I\kern-.18em P}}
\def\IR{\relax{\rm I\kern-.18em R}}
\def\IN{\relax{\rm I\kern-.18em N}}
\def\IZ{\relax\ifmmode\mathchoice
{\hbox{Z\kern-.4em Z}}{\hbox{Z\kern-.4em Z}}
{\lower.9pt\hbox{Z\kern-.4em Z}}
{\lower1.2pt\hbox{Z\kern-.4em Z}}\else{Z\kern-.4em Z}\fi}

\def\SL{${\rm SL}(2,\IR)$}

\def\zbar{\bar z}

\catcode`\@=12

\Title{RU-91-51}{An Introduction to 2d Gravity
and Solvable String Models\abstractfont\footnote{$^+$}{Lectures at the 1991
Trieste Spring School.}}
\centerline{Emil J. Martinec\footnote{$^\dagger$}
{Research supported in part by DOE contract DE-FG02-90ER-40560,
an NSF Presidential Young Investigator Award, and the Alfred P. Sloan
Foundation.}}
\bigskip\centerline{Enrico Fermi Institute and
Department of Physics\footnote{$^*$}{Permanent address}}
\centerline{University of Chicago,Chicago IL 60637}
\bigskip\centerline{{\it and}}
\bigskip\centerline{Dept. of Physics and Astronomy, Rutgers University}
\centerline{Piscataway, NJ 08854}
\vskip .5in

Continuum and discrete approaches to 2d gravity coupled to
$c<1$ matter are reviewed.

\Date{09/90}

\bigskip\centerline{{\titlerm An Introduction to 2d Gravity}}
\centerline{{\titlerm and Solvable String Models}}
\bigskip\centerline{E. Martinec}
\centerline{Enrico Fermi Institute and
Department of Physics\footnote{$^*$}{Permanent address}}
\centerline{University of Chicago,Chicago IL 60637}
\centerline{{\it and}}
\centerline{Dept. of Physics and Astronomy, Rutgers University}
\centerline{Piscataway, NJ 08854}

\newsec{Introduction}

Two-dimensional gravity has undergone a thorough examination
over the last few years, especially with the emergence of efficient
calculational techniques stemming from the matrix model approach.
Two dimensions is an arena where rather complex phenomena
(confinement, chiral symmetry breaking, integrability, nonperturbaive
phenomena, solitons; the list is extensive) can be stripped
of complications of higher dimensional kinematics and
dynamics while hopefully retaining many of the physical features
of the problem of interest.
Thus we study two dimensional gravity as a model for exploring
the structure and formalism of quantum gravity: the wavefunction
of the universe, the Wheeler-DeWitt equation\ref\hh{J. Hartle and
S. Hawking\journal Phys. Rev.&D28 (83) 2960;
B.S. DeWitt\journal Phys. Rev.&160 (67) 1113.},
the meaning
of measurement in quantum gravity, the statistical properties of
the metric and matter in a fluctuating geometry, etc.
Indeed the collection of solvable 2d gravity-matter systems
provides a rich laboratory for the investigation of these issues.
In addition, since unified strings are by definition coordinate
invariant 2d quantum field theories these systems are equally well
regarded as solvable models of string theory, where
one might begin to formulate a useful string field theory\ref\dj{S.
Das and A. Jevicki\journal Mod. Phys. Lett.&A5 (90) 1639.}, study
nonperturbative effects\ref\shenker{S. Shenker, Rutgers preprint
RU-90-47, to appear in the proceedings of the Cargese workshop (1990).},
strong coupling phenomena, etc.

There are several ways one might approach the continuum theory
of 2d gravity.  One is to write down a field theory on 2d metrics
and matter\ref\pol{A.M. Poyakov\journal Phys. Lett.&103B (81) 207.},
regulate it covariantly, and try to find a consistent
renormalization to the continuum limit\ref\ct{T. Curtright and
C. Thorn\journal Phys. Rev. Lett.&48 (82) 1309; \andjournal
Ann. Phys.&147 (83) 365;
\andjournal Ann. Phys.&153 (84) 147}\ref\gn{J.-L. Gervais and
A. Neveu\journal Nucl. Phys.&B199 (82) 59;
{\bf B209} (1982) 125; {\bf 224} (1983) 329; {\bf 238} (1984) 125, 396;
\andjournal Phys. Lett.& 151B (85) 271.; \nextline
J.-L. Gervais\journal Phys. Lett.& 243B (90) 85; \andjournal
Comm. Math. Phys.&130 (90) 257.}.  Another is to
discretize the theory and study the fluctuations of the discrete
geometry\ref\randsurf{F. David\journal Nucl. Phys.&B257[FS14] (85) 433;
V. Kazakov\journal Phys. Lett.&119A (86) 140; {\bf 150B} (1985) 282;
V. Kazakov, I. Kostov, and A.A. Migdal\journal Phys. Lett.&B157 (85) 295;
J. Ambjorn, B. Durhuus, and J. Frohlich\journal Nucl. Phys.&B257 (85) 433.},
then try to take the continuum limit
of the random lattice model (a method that has also been
used for strings embedded in higher dimensional
spacetimes\ref\regge{I. Klebanov and L. Susskind\journal Nucl.
Phys.&B309 (88) 175.}).
Both should yield the same results if indeed 2d gravity is universal,
\ie\ if there is a second order phase transition for
the fluctuating surfaces in some region of the space of
coupling constants (cosmological constant, topological coupling,
boundary cosmological constant, etc.).
The former line of attack adheres more closely to the standard
conceptual framework of gravity, yet in two dimensions the latter
formulation has yielded more success to date.  This success
comes about largely because
in two dimensions it is not difficult to enumerate the
lattice configurations, which are gauge invariant.

These lecture notes introduce the present understanding
of 2d gravity/solvable string models, beginning in section 2
with a review of the continuum formulation in terms
of the functional integral over metrics.  While there is
little that can be computed exactly, we argue that much
of the classical structure should persist upon quantization;
and that the familiar structures of rational conformal field theory
might appear, albeit in a much more subtle and complicated way.
In section 3 we discuss the behavior of the theory
under rescalings of the 2d metric\ref\kpz{V. Knizhnik,
A.M. Poyakov, and A.B. Zamolodchikov\journal Mod. Phys. Lett.&A3 (88) 819;
F. David\journal Mod. Phys. Lett.&A3 (88) 1651; J. Distler and
H. Kawai\journal Nucl. Phys.&B321 (89) 509.}
as well as the related
dynamics in the zero mode sector\ct\ref\seiberg{N. Seiberg, {\it Notes
on Liouville Theory and Quantum Gravity}\journal Prog. Theor. Phys.,
Suppl.&102 (90) 319.}\ref\joe{J. Polchinski,
`Remarks on the Liouville field theory', Texas preprint UTTG-19-90,
 presented at Strings '90, College Station, 1990.}
\ref\mss{G. Moore, N. Seiberg, and M. Staudacher\journal
Nucl. Phys.&362 (91) 665; G. Moore and N. Seiberg, Rutgers
preprint RU-91-29.}.
In section 4 we switch to the discrete formulation.  Here
the partition function of random surfaces can be recast as
and integral over N$\times$N matrices\randsurf\ where in the continuum
limit $N\rightarrow\infty$\ref\bkdsgm{E. Br\'ezin and V. Kazakov\journal
Phys. Lett.&236B (90) 14; M. Douglas and S. Shenker\journal Nucl. Phys.&B335
(90) 635; D. Gross and A.A. Migdal\journal Phys. Rev. Lett.&64 (90) 127;
\andjournal Nucl. Phys.&340 (90) 333.}.
Both the lattice\ref\discrete{L. Alvarez-Gaum\'e, C. Gomez, and
J. Lacki\journal Phys. Lett.&253B (91) 56; A. Gerasimov, A. Marshakov,
A. Mironov, A. Morozov, and A. Orlov\journal Nucl. Phys.&B357 (91) 565;
E. Martinec\journal Comm. Math. Phys.&138 (91) 437.}
and continuum\bkdsgm\ref\bdss{T. Banks, M. Douglas, N. Seiberg,
and S. Shenker\journal Phys. Lett.&238B (90) 279.}\ref\douglas{M.
Douglas\journal Phys. Lett.&238B (90) 176.} theories exhibit a rich
integrability structure.  The scaling behavior predicted by the
continuum approach is recovered as well as the zero mode
dynamics\seiberg\mss.

\newsec{The effective action for gravity and its quantisation}

A crucial feature of 2d quantum gravity is that it must
be scale invariant as well as reparametrization invariant\ref\friedan{D.
Friedan, in the proceedings of the 1982 Les Houches Summer
School; and unpublished.};
local shifts $g_{ab}\rightarrow e^\eps g_{ab}$
of the metric scale factor are simply shifts of an
integration variable in the functional integral over metrics.
The effect of such a shift is to produce the trace of the
stress tensor, which
we can generally expand in terms
of scaling fields
 \eqn\weyl{\vev{T_{z\zbar}}=\frac\delta{\delta\eps(z)}\ZZ=\int e^{-S}
	\beta_i\OO^i\ .}
The $\beta_i$ are the beta functions of the theory.
The result \weyl\ must vanish apart from possible contact terms
with local operators.
Often we are interested in gravity coupled to some number $d$ of
scalar matter fields
  \eqn\pertaction{ S=\coeff1{4\pi}\int \sqrt g
	[T(X)+R^{\sst(2)}D(X)+\p_aX^\mu\p_b X^\nu
		(G_{\mu\nu}(X)g^{ab}+B_{\mu\nu}(X)\eps^{ab})]\ . }
in which case the vanishing trace of the stress tensor
enforces a set of conditions on the spacetime fields which are
the equations of motion of string theory.  One may adopt either
of two points of view: In `critical' string theory, one sets
$\frac\delta{\delta\eps}\ZZ_{\sst matter}=0$; then one has
an additional symmetry -- Weyl invariance -- which one can
quotient by, so that one integrates over the gravitational
measure $\frac{\DD g_{ab}\DD X}{{\rm Diff}\sdtimes{\rm Weyl}}$.
In this way one regards all components of the 2d metric as gauge
degrees of freedom.
On the other hand, from the `noncritical' string point of
view, one can trivially achieve scale invariance by integrating
over all possible scale factors
$\frac{\DD g_{ab}\DD X}{{\rm Diff}}$.
This is equivalent to a particular class of solutions to critical
string theory if we regard the local scale factor as another
scalar field like the $X$'s\ref\lioutime{F. David and E. Guitter\journal
Europhys. Lett.&3 (87) 1169; \andjournal Nucl. Phys.&B293 (88) 332;
S.R. Das, S. Naik, and S. Wadia\journal Mod. Phys. Lett.&A4 (89) 1033;
J. Polchinski\journal Nucl. Phys.&B324 (89) 123;
T. Banks and J. Lykken\journal Nucl. Phys.&B331 (90) 173.};
perhaps it is completely equivalent
if we complexify the space of metrics.  In either case the
object is to compute and solve the equations of
vanishing stress tensor.
There are two common methods of calculation: the spacetime weak field
expansion\ref\smatrix{C. Lovelace\journal Nucl. Phys.&B273 (86) 413.}
and the 2d loop expansion\ref\sigmodel{D. Friedan\journal
Phys. Rev. Lett.&45 (80)
1057; C. Callan, D. Friedan, E. Martinec,
and M. Perry\journal Nucl. Phys.&B262 (85) 593;
E.S. Fradkin and A. Tseytlin\journal Phys. Lett.&158B (85) 316;
\andjournal Nucl. Phys.&B272 (86) 647.}.  In the former
one expands around a known fixed point (\eg\ free field theory in
$d$ spacetime dimensions)
  $$ S=\coeff1{4\pi}\int \p X\bar\p X+
	\delta T(X)+\p X^\mu\bar\p X^\nu
	\bigl(G_{\mu\nu}\delta D(X)+\delta G_{\mu\nu}(X)
		+\delta B_{\mu\nu}(X)\bigr)\ .$$
Bringing down powers of the composite operator perturbations
will reliably find nearby fixed points (unless one has chosen a singular
parametrization of the coupling space).  The origin of the
beta function is the overall scale divergence
  $$\int \frac{d^2\lam}{|\lam|^2}\lam^{L_0+\bar L_0}\int_{\sst N-1\ z's}
	\vev{\prod_{i=1}^N\OO_i(\lam z)} $$
The similarity of this expression to the Koba-Nielsen
prescription for the calculation of the string S-matrix
is not coincidental\ref\polch{J. Hughes, J. Liu, and
J. Polchinski\journal Nucl. Phys.&B316 (89) 15.}.
Indeed, the effective action
of string theory (actually any field theory)
is determined from the S-matrix by subtracting the contributions
of intermediate on-shell poles.  In the Koba-Nielsen formula
these are the logarithmic subdivergences in the integrations
over the locations of composite operators; the
beta functions are the equations of motion following from
this effective action.  Note that the kinetic operator
$\p^2 S/\p g_i\p g_j=\p\beta_i/\p g_j$
for small fluctuations is the anomalous dimension operator --
the linearized scaling operator $L_0+\bar L_0-2$.
The disadvantage of the spacetime weak field expansion is that
one must work in a particular coordinate basis in 2d field
space, so spacetime general coordinate invariance (invariance
under 2d field redefinitions of the $X$'s) is not manifest.
This is an advantage of the 2d loop expansion; rather than
working in a specific basis of tensor fields on the spacetime
manifold, one expands in small fluctuations of the coordinates
$X$, which can be made manifestly generally covariant\sigmodel.
This perturbation series is an expansion in spacetime variation
of the string coordinates relative to the 2d Planck constant,
the inverse string tension $\alpha'$.  A superposition of these
two methods yields the beta functions\sigmodel
  \eqn\betafn{\eqalign{\beta^G_{\mu\nu}=&R_{\mu\nu}-2\nabla_\mu\nabla_\nu D
	+\nabla_\mu T\nabla_\nu T=0\cr
	\beta^D=&\frac{26-d}{3\alpha'}+R+4(\nabla D)^2
	-4\nabla^2 D+(\nabla T)^2+V(T)=0\cr
	\beta^T=&-2\nabla^2T+4\nabla D\nabla T + V'(T)=0}  }
in a double expansion in field strength and $\alpha'$; $V(T)$
is a generic potential $V(T)=\hf T^2+ O(T^3)$.
Both expansions are required here; the loop expansion for
general covariance, and the weak field expansion to incorporate
the tachyon.

{}From the critical string viewpoint, we are interested in
strings in $d$ spacetime dimensions.
The equations \betafn\ are solved by ($\phi=X^0$, say)
  \eqn\soln{\eqalign{ T=&\coeff{\mu}{2\gamma^2}e^{\gamma\phi}\qquad,\qquad
		Q=\coeff 2\gamma+\gamma=\sqrt{\coeff{26-d}3}\cr
	D=&\coeff 1\gamma\phi\cr
	G_{\mu\nu}=&\delta_{\mu\nu}\ ;}}
Although each equation of motion in \betafn\ is satisfied at its
leading nontrivial order in powers of $T$, none is solved at subleading
order.
However we are only considering the lowest order equations; one
might hope that higher orders correct the problem.
After all, the $\alpha'$ (loop) expansion is not valid here since
$\gamma$ is not small; the kinetic term in the tachyon
beta function is only found after a resummation of loops,
but then we have no reason to ignore terms involving the gradient
of the dilaton field.
Fortunately we are looking for a solution with rather special
properties: it is \soln\ at lowest order and its
renormalization involves only $\phi$-dependent fields.
Since $\nabla_\phi T\propto T$, the exact solutions $\widehat G$,
$\widehat D$, $\widehat T$ are power series in $T$ (assuming the
weak field expansion is summable).  Therefore we can find a field
redefinition -- the reversion of the power series $\widehat G(T)$,
$\widehat D(T)$, $\widehat T(T)$ -- such that \soln\ is the
{\it exact} solution\foot{This argument is due to T. Banks.}.
The importance of $\nabla T\propto T$ is that the field redefinition
required is local in spacetime.

The noncritical string viewpoint is somewhat more helpful in this situation.
Here we have $d-1$ string coordinates coupled to a dynamical metric
$g_{ab}=e^{\gamma\phi}\hat g_{ab}$ where $\hat g_{ab}$ is
a fixed metric (up to moduli).  For $d-1$ free scalar fields we have
  $$ \beta=\coeff{26-d}{3\alpha'}\sqrt gR^{\sst(2)}+
	\coeff\mu{2\gamma}\sqrt{g}  $$
Integrating $\delta S_{eff}/\delta\phi=\beta$ gives an effective classical
action, the Liouville action
  \eqn\Sliouv{S_{eff}=\frac1{4\pi}\int\hf(\p\phi)^2
	+\coeff1\gamma\phi R^{\sst(2)}+
	\coeff\mu{2\gamma^2}e^{\gamma\phi} }
for the dynamics of the metric.  The integral over metrics
restores scale invariance if we can succeed in correctly performing
the functional integral or otherwise quantizing the theory.
{}From critical string considerations, one might expect a problem
since the tachyon stress-energy appears
to feed into the metric  beta function.  There is
in fact a renormalization prescription such that $S_{eff}$
is conformally invariant\friedan.  Since the metric is
dynamical, the coordinate regularization scale and the covariant
scale are related by a fluctuating field: $\delta=e^{\gamma\phi/2}\eps$.
Therefore a regulated action should be, \eg
  $$S^{\sst Reg}_{eff}=\frac1{8\pi}\int d^2z\;\p\phi
	\bigl(e^{\eps^2e^{-\gamma\phi}\p^2}\bigr)\bar\p\phi+QR^{\sst (2)}
	\phi+\coeff{\mu^2}{\gamma^2}e^{\gamma\phi}\ ,  $$
and we should calculate $\delta S_{eff}/\delta\phi=\beta$ incorporating
the effects of the regulator.
In the loop expansion, normal ordering an exponential means resumming
self-contractions; this introduces cutoff-dependence, and therefore
$\phi$-dependence.
One finds that the entire effect of imposing
$\beta_{eff}=0$ is to renormalize the parameters in the
Liouville lagrangian: $\mu\rightarrow\mu_{ren}$ (which can be absorbed in
a shift of $\phi$), and $Q\rightarrow Q_{ren}=\frac2\gamma+\gamma$\ct\friedan.
Essentially we are allowed to ``normal order'' the Liouville term,
and provided its quantum scale dimension is (1,1) and $\gamma\le\frac Q2$
all is well.
Physically, at short distances ($\phi\rightarrow-\infty$)
the exponential potential term is small, so divergences should be those
of free field theory.
It seems that the complicated field redefinition of the critical
string theory amounts to the change from coordinate to covariant
regulator on the world sheet;
recall that spacetime field redefinitions are equivalent
to a change of renormalization scheme in the 2d sigma model.
It is intriguing that in this low-dimensional context there can
be nontrivial solutions of the string equations of motion
whose stress-energy does not
back-react on the spacetime metric.

One indication of the conformal invariance of quantum Liouville theory
is the degree of singularity of the operator product expansion of
the Liouville perturbation; assuming free field operator products
(but see \joe\seiberg)
  \eqn\expope{e^{\gamma\phi(z)}\; e^{\gamma\phi(w)}\sim|z-w|^{-2\gamma^2}
	e^{2\gamma\phi(z)}+\ldots\qquad.}
In the weak coupling regime $\gamma\ll 1$ the singularity
is integrable and the beta function calculated above will continue
to vanish at first order in the perturbed theory.  This reasoning
does not give the correct upper bound on $\gamma$ however,
indicating that problems set in for $\gamma\ge 1$ ($c\ge -2$) rather
than the observed upper bound $\gamma=\sqrt 2$ ($c=1$).
Ordinarily in conformal field theory, an operator that produces
the identity in its operator product with itself has
  $$\OO(z)\OO(w)\sim|z-w|^{-4h}\One$$
from which one can read off the dimension of that operator.
It seems to be that, even though it does not produce the
identity, in its self-product the singularity of $e^{\gamma\phi}$
is four times its `effective dimension'.
Then from \expope\ we would conclude that \ephi\ becomes
effectively irrelevant at $c=1$, or more generally $\OO e^{\alpha\phi}$
becomes effectively irrelevant for $\alpha> Q/2$.
This behavior is rather similar to what happens when attempting to
incorporate massive string states in the sigma model lagrangian
of the critical string.  The on-shell vertex operator for such a perturbation
has $h=1$, but only by virtue of an irrelevant
spatial operator coupled to a negative dimension
temporal operator.  The disease of irrelevant operators, namely
nonrenormalizability, shows up in the appearance of many low
dimension operators with highly singular coefficients in the
self-OPE.  Such a similar explosion occurs for
operators $\OO e^{\alpha\phi}$ with $\alpha> Q/2$ (for instance
the `black hole perturbation'\ref\blah{E. Witten, IAS preprint
IASSNS-HEP-91/12; G. Mandal, A. Sengupta, and S. Wadia,
IAS preprint IASSNS-HEP-91/10.} $\p X\bar\p X e^{Q\phi}$,
hence it is in no sense a `perturbation' of flat spacetime
Liouville theory).

To recapitulate, we have
  $$\eqalign{T_{z\zbar}=&0\qquad ({\rm Liouville\ e.o.m.})\cr
	T_{zz}=&-\hf(\p\phi)^2+\coeff Q2\p^2\phi\qquad Q=\coeff 2\gamma
		+\gamma\cr
	h_{\exp[\alpha\phi]}=&-\hf\alpha(\alpha-Q)\cr}$$
These look like free field results, but it must be stressed that
$\vev{\phi\phi}$ is {\it not} the free field propagator.
The Liouville equation is geometrically the equation for constant
(negative for $\mu>0$) curvature surfaces $R=-\mu$.  The semiclassical
expansion is valid for $\gamma<<1$; rescaling $\phi\rightarrow
\frac 2\gamma\phi$ puts an overall factor of $1/\gamma^2$ in
front of \Sliouv.  If we plot
the potential for the zero mode

\vbox to3in{\vfil}

\noindent we see that a stable solution exists only for surfaces
of genus $g>1$ (in the absence of point sources of curvature;
see below).  The stable point $\vev\phi$
increases with genus; in fact there is a scaling relation\kpz:
let $\phi\rightarrow\phi+\frac1\gamma\log a$, then
  $$\ZZ(g_{str},\mu)=\ZZ(g_{str},a\mu)a^{-(2-2g)\frac Q{2\gamma}}\ ,  $$
implying
  \eqn\kpzscal{ \ZZ_g(g_{str},\mu)=C_g\ \mu^{(2-2g)\frac Q{2\gamma}}\ . }

The full classical solution to the Liouville equation of motion
is discovered through its geometrical role; $e^{\gamma\phi}$
must be a density under coordinate transformations,
$e^{\gamma\phi(z)}=|\frac{\p z'}{\p z}|^2e^{\gamma\phi(z')}$.
Since the standard constant negative curvature metric on the upper
half plane (UHP) is $ds^2=dzd\zbar/({\rm Im}\,z)^2$, $\phi$ must
look locally like
  $$ \phi=\frac1\gamma\log\left[\frac{16}\mu
	\frac{\p A(z)\;\bar\p A^*(\zbar)}{(A-A^*)^2}\right]\ , $$
where $A$, $A^*$ are local coordinates on the surface; \ie\
$A(z)$ is the map from the UHP to the Riemann surface $\Sigma$
  $$\coeff\mu{16}e^{\gamma\phi}dzd\zbar=\frac{dAdA^*}{(A-A^*)^2}\ .$$
The line element on the UHP is invariant under SL(2,R)
transformations $A\rightarrow\frac{aA+b}{cA+d}\equiv g(A)$.
This transformation must leave $\phi$ invariant, but may do so
in a nontrivial way, \eg\ by making a circuit of a nontrivial
closed path on $\Sigma$ (see fig.2).  That is, the {\it monodromy}
of the local coordinate $A$ is a set of \SL\ transformations
which cover the surface $\Sigma$ onto the UHP.  There are several
classes of monodromies:

\vbox to2.5in{\vfil}

\item{(1)} elliptic monodromy -- $g=\left({a\atop c}{ b\atop d}\right)$
is conjugate to a rotation, \ie\
there exists $h$ such that $hgh^{-1}=\left({\cos\theta\atop\sin\theta}{
-\sin\theta\atop\cos\theta}\right)$.  The surface $\Sigma$ has a conical
singularity of deficit angle $\theta$.  Deficit angles
$\theta=\frac{2\pi}j$, $j\in\IZ$, are `nice' since
they are covered by the UHP;
$j=\frac mn$ requires an $n$-fold branched cover of the UHP at
the fixed point of the rotation.
\item{(2)} parabolic monodromy -- $h$ exists such that $hgh^{-1}
=\left({1\atop 0}{u\atop 1}\right)$, a translation
(in terms of deficit angles,
$\theta=\pi$ and the surface has a cusp).
\item{(3)} hyperbolic monodromy --
$hgh^{-1}=\left({\lam\atop0}{ 0\atop\lam^{-1}}\right)$
is a dilation.  The identification of the UHP under $g$ makes
a handle.

A complete solution of the Liouville equation consists of representing
the surface $\Sigma$ by its fundamental group, the discrete subgroup
(Fuchsian group) of \SL\ that covers the surface onto the UHP; then
one constructs the automorphic function $A(z)$ convariant under
this group.

The three types of monodromy above are continuously related, as can be seen
by pinching a handle on $\Sigma$:

\vbox to1.5in{\vfil}

\noindent In the classical theory,
a deficit angle $\theta$ is a delta-function
source of curvature:
  \eqn\Leom{\coeff1{4\pi}\p^2\phi+\coeff\mu{8\pi\gamma}e^{\gamma\phi}
	=\coeff{Q}{8\pi}\hat R
	=\sum_i\coeff{\theta_i}{\pi\gamma}\delta^{\sst (2)}(z-z_i)\ .  }
Integrating both sides, one sees that there is a solution whenever
$2-2g+\theta_i/\gamma<0$.
Eq. \Leom\ is the saddle point of the functional integral
  \eqn\Liouv{\int\DD\phi\;e^{-S_{eff}}\prod_i
	e^{\frac{\theta_i}{\pi\gamma}\phi(z_i)}\ .  }
On the other hand, there is no local source for hyperbolic
monodromy (the fixed points of the \SL\ transformation $g$ on
the UHP are not on the surface $\Sigma$).  In terms of Liouville dynamics,
the initial field configuration for $\phi$ (along some closed loop
generating the monodromy)
never propagates to $\phi=-\infty$
where the field configuration can be interpreted as localized at a point
in the covariant metric.
Note also that there is no classical geometrical interpretation for
deficit angle $\theta>\pi$;
\ie\ when the source contributes half the curvature contribution
$\frac1{8\pi}Q\int R$ of the sphere in the functional integral.
Two parabolic cusps turn a sphere into
a cylinder; we cannot go beyond this while maintaining the geometrical
interpretation of $\phi$.
Quantum mechanically this means\seiberg\
$\alpha=\frac\theta{\pi\gamma}\le\frac Q2=\frac1\gamma+\frac\gamma2$.
This does not mean that operators with such Liouville
charge `don't exist', rather merely
that we cannot give them a geometrical interpretation.
{}From the discussion of section (2), we conclude that gravitationally
dressed operators with $\alpha> Q/2$ are `effectively irrelevant'.
If we perturb the action by them,
new dimension one operators will have to be added to the action to
subtract singularities, a procedure that rapidly snowballs; as vertex
operators they cannot be renormalized simply by normal ordering
because the loop expansion is not well-behaved.
It is doubtful that KPZ scaling (see below) can be maintained.

The above considerations motivate a brief review of \SL\
representation theory\ref\sltwo{see, \eg, A.O. Barut and C. Fronsdal\journal
Proc. Royal Soc.&A287 (65) 532.}.  Representations are labelled by their
values of the quadratic Casimir
$C_2=j(j-1)$ and $J^3=m+E_0$, $m\in \IZ$.
All representations can {\it formally} be built from the two-dimensional
representation
$\left({w_1\atop w_2}\right)\rightarrow\left({g(w_1)\atop g(w_2)}\right)=
\left({a\atop c}{b\atop d}\right)\left({w_1\atop w_2}\right)$.
Realizing the \SL\ algebra
on differential operators $J^+=\frac1{\sqrt2}w_1\p_2$,
$J^-=\frac1{\sqrt2}w_2\p_1$, $J^3=\hf(w_1\p_2-w_2\p_1)$,
the monomial $N_mw_1^aw_2^b=N_m(w_1w_2)^j(w_1/w_2)^{E_0+m}$
transforms as part of the spin $j$ representation, where $N_m$ is
a normalization.  There are several cases:

\item{(1)} The trivial representation $j=E_0=m=0$.

\item{(2)} The finite dimensional representations $-2j\in\IN$, $E_0=0$,
$m=-j,...,j$; these are not unitary.

\item{(3)} Discrete series representations $\DD_\pm$: For $\DD_+$, we
have $-j+E_0=0$, $-2j\not\in \IN$, $J_3+E_0=0,-1,-2,...$; for $\DD_-$,
$j+E_0=0$, $-2j\not\in\IN$, $J_3+E_0=0,1,2,...$.  These are unitary
if $E_0\in\IR$ and $j>0$.  They are related to the elliptic monodromy
conjugacy classes of \SL\ and have $C_2>0$.

\item{(4)} Continuous series representations: $-\hf<{\rm Re}\,E_0\le\hf$,
$-j+E_0\ne0,\pm1,\pm2,...$, $J_3+E_0=0,\pm1,\pm2,...$.  These
are unitary if $E_0\in\IR$, $j-\hf\in i\IR$; or $E_0,j\in\IR$,
$|-j+\hf|<\hf-|E_0|$.  These are related to hyperbolic conjugacy
classes in \SL\ and have $C_2<0$.

How are these representations related to Liouville theory?
Recall that classically $T_{zz}=-\hf(\p\phi)^2+\frac1\gamma\p^2\phi$.
{}From this follows\gn\ref\tak{P. Zogrof and
L. Takhtadjan\journal Math. USSR Sbornik&60 (88) 143.}
  \eqn\Teqn{(\p^2+\coeff{\gamma^2}2 T(z))e^{-\frac\gamma2\phi}=0\ ,  }
and from the classical solution we have
  \eqn\classical{\eqalign{e^{-\frac\gamma2\phi_{cl}}=&\sqrt{\frac{16}\mu}
	\frac{(A-A^*)}{\sqrt{\p A}\sqrt{\bar\p A^*}}\cr
	=&\sqrt{\frac{16}\mu}\sum_{m=1,2}\psi_{j,m}(z)
		\psi^*_{j,m}(\zbar)}  }
for $j=-\hf$, where
  $$w_1\equiv\psi_{-\frac12,\frac12}=\frac1{\sqrt{\p A}}\quad;\qquad
	w_2\equiv\psi_{-\frac12,-\frac12}=\frac A{\sqrt{\p A}}\ .  $$
are the two solutions to \Teqn.
The expressions \classical\ can be regarded as a classical version of
conformal field theory where each measurement is a sum of
holomorphic times antiholomorphic `chiral vertices',
glued together in a monodromy invariant
way to make physical single-valued expressions.  From the
two basic solutions one can in principal build
all exponentials corresponding to finite-dimensional
representations, \eg\
  $$e^{j\gamma\phi}=
	\sqrt{\frac{16}\mu}\sum_{m=-j,...,j}\psi_{j,m}(z)
		\psi^*_{j,m}(\zbar)\quad,\qquad-2j\in\IN  $$
An important issue is whether such expressions generalize to other
classes of representations for which the sum over $m$ is
infinite.
In principle one should be able to extract the answer
from the classical solution $A(z)$ to \Leom, the saddle point
of the correlation \Liouv; this is in turn
determined by its monodromy $\Gamma\in{\rm SL(2,\IR)}$.

The quantum theory may have a structure similar to
that of known conformal quantum field theories: the correlation
functions would be (generically infinite) sums of holomorphic
times antiholomorphic conformal blocks glued together to make
a monodromy invariant object\ref\bpz{A. Belavin, A. Polyakov,
and A.B. Zamolodchikov\journal Nucl. Phys.&B241 (84) 333.}
  \eqn\correl{\vev{\prod \OO_i(z,\zbar)}
	=\sum_\alpha \FF_\alpha(z)\FF^*_\alpha(\zbar)\ . }
For instance, in current algebra
conformal field theories the conformal blocks $\FF_\alpha$ carry
two sets of indices, an `external' index (like $i$ in \correl)
which is a representation
label for the current algebra, and an `internal' index which is
a quantum group index, since the monodromy acts on the blocks
like the $R$-matrix of a quantum group\ref\alekshat{A. Alekseev and
S. Shatashvili\journal Comm. Math. Phys.&133 (90) 353.}.
In the case of Liouville, there is no
`external' symmetry group other than Virasoro, and the index is
continuous; however there
is an `internal' symmetry group $SL_q(2)$ which is, in some sense
yet to be understood, the `quantization' of the classical
monodromy action on $A(z)$\gn\tak.  In this respect Liouville is like
other coset conformal field theories (Liouville can be thought
of as the coset conformal field theory based on
$SL(2,R)/N$, where $N$ is the Borel subgroup\ref\borel{A. Polyakov\journal
Mod. Phys. Lett.&A2 (87) 893; V. Kniznik, A. Polyakov, and
A.B. Zamolodchikov\journal Mod. Phys. Lett.&A3 (88) 819; A. Alekseev
and S. Shatashvili\journal Nucl. Phys.&B323 (89) 719;
M. Bershadsky and H. Ooguri\journal Comm. Math. Phys.&126 (89) 49.}),
where the external symmetry is gauged and the internal quantum
group symmetry remains but is `confined' -- only appearing when
one pulls the theory apart into holomorphic constituents.
In terms of the parameter $\gamma$ in the Liouville lagrangian,
the quantum group parameter is  $q=\exp[i\pi\gamma^2]$\ref\tak{F. Smirnov
and L. Takhtadjan, U. Col. Boulder preprint \#12 in
Applied Mathematics (1990).}\gn.
The translation into \SL\ notation of the condition $\alpha\le Q/2$
means that $e^{j\gamma\phi}$ must have $j\le \frac Q{2\gamma}
=\frac1{\gamma^2}+\hf$; the quantum dimension is
$h_j=-j-\frac{\gamma^2}2 j(j-1)$.  In these expressions the first
term is the classical value; the second is the quantum correction.
Note that $L_0$ is also real for $j=\frac Q{2\gamma}+i\lam$,
$\lam\in\IR$, for which
$h_j=\frac{\gamma^2}2[(\coeff{Q}{2\gamma})^2+\lam^2]$.

An analogy with the SU(2) WZW theory might be helpful here.
The classical solution to the theory is
  $$g_{ab}(z,\zbar)=\sum_c g_{ac}(z)\bar g_{cb}(\zbar)$$
where the holomorphic part $g(z)$ transforms under left multiplication
by the loop group of SU(2) and the Virasoro algebra, and also by right
multiplication under SU(2).  The holomorphic constituents
are determined by their monodromy $g(z)\rightarrow g(z)\cdot h$
around nontrivial closed loops on $\Sigma$.
$g_{ab}(z)$ intertwines with $g_{cd}(z')$ via a classical
$r$-matrix; upon quantization this internal index labels
an $SU_q(2)$ quantum group representation with
$q=\exp(\frac{2\pi i}{k+2})$\alekshat, and the intertwining is via
a quantum $R$-matrix.
For SL(2) and for Liouville a major complication is that the
fusion of representations closes on the continuous series
representations and it is not clear that one has a useful
or effective way of decomposing the correlations on intermediate states.

Although the above analysis is rather appealing as it places the
Liouville theory within the customary conceptual framework
of conformal field theory, it has not proven to be sufficiently
powerful to enable the calculation of correlation functions.
It is a useful route in the case of rational conformal field theories
because the monodromy representations are finite, related to a
closed system of differential equations that one may derive
from various Ward identities\bpz.
In the case of Liouville theory,
and also noncompact current algebra, the monodromy representations
are infinite-dimensional; and the Ward identities are not
sufficiently powerful to give a closed system of differential
equations that will determine the correlation functions.  Although
finite dimensional representations $e^{-j\gamma\phi(z)}$ lead
to finite order differential equations in $z$,
general matter operators require infinite dimensional representations
which don't satisfy any simple equations; such representations
are required in intermediate states and therefore upon
factorization, hence the dependence of correlations on the
position of these operators is an open question.
It seems
that at least in the case of Liouville theory, other analytic
techniques are available
(see the lectures of D. Kutasov at this school, and references therein).

The above analysis points to a picture of `quantum Riemann surfaces'
where the classical solution $A(z)$ of Liouville theory is deformed
into some kind of quantum conformal block, its classical \SL\
monodromy deforming into an $\rm SL_q(2)$ quantum group structure
with $q$ related to the coupling
constant $\gamma$ of Liouville theory.

\newsec{Scaly behavior}

Regardless of our understanding of how to compute Liouville
correlation functions, there are several general statements
we can make about those correlations {\it assuming} the
quantization preserves free-field ultraviolet behavior even
in the presence of the exponential interaction term.  The
strongest `theoretical' reason for such a presumption (made implicitly
by the authors of \kpz\ is that physical short distance
as measured in the metric $ds^2=e^{\gamma\phi}dz^2$ is where the
Liouville potential is exponentially small, and so should not affect
the free field results.  The `experimental' evidence for the
validity of this assumption is the agreement of predicted
scaling relations\kpz, and more recently tree-level S-matrix elements,
with the continuum limit
of discretized random surfaces in the matrix model\randsurf.

What are the Liouville predictions?  Suppose we wish to study
conformal matter with central charge $c$($=d$ above) coupled to
2d gravity.  The matter theory will contain a set of scaling operators
$\OO_k^{\sst matter}$ of scale dimension $h_k$.
A generally covariant theory must
make $\OO_k^{\sst matter}$ into a coordinate density of scale
dimension $(h^{\sst tot},\bar h^{\sst tot})=(1,1)$; for instance,
  \eqn\dressing{\OO_k^{\sst grav}=e^{\alpha_k\phi}\OO_k^{\sst matter}}
with
  \eqn\mashl{h_{\sst total}=-\hf\alpha_k(\alpha_k-Q)+h_k=1\ .}
Then the integrated
correlation functions of $\OO_k^{\sst grav}$ are 2d general
coordinate invariant.
The exponential Liouville dependence of such operators will
modify KPZ scaling in correlation functions, cf \Liouv.
The string theory interpretation
of this {\it gravitational dressing} is that the Liouville field $\phi$
is the (Euclidean) time component of the string's position in
spacetime; the time ($\phi$) dependence of \dressing\
is just that of a linearized
solution of the (Euclidean) spacetime equations of motion with
energy $\alpha_k=j_k\gamma$; and the relation
$-\hf(\alpha_k-Q)^2+h_k=\frac{1-d}{24}$
between $\alpha_k$ and $h_k$
is the mass shell condition for linear perturbations, \ie\
$h_k$ is the eigenvalue of the spatial Laplacian $L_0^{\sst matter}$.
This makes it clear that conformal matter is from the
spacetime point of view the special class of stationary solutions
of the string equations.
{}From the previous analysis of Liouville theory, if
$1-h_k>\frac{Q^2}8$ then there does {\it not} exist a {\it local}
generally covariant measurement corresponding to the matter
operator $\OO_k$; \ie\ $\OO_k^{\sst matter}$ looks local in
coordinates $(z,\zbar)$, but its gravitational dressing
requires hyperbolic monodromy $j_k=\coeff Q{2\gamma}+i\lam$ which has no local
interpretation in the covariant theory.  Seiberg\seiberg\ has dubbed
such operators tachyonic since their spacetime mass shell
condition \mashl\ implies imaginary mass for the corresponding string
state in spacetime.  The reason that some perfectly sensible
local operators may not have a local gravitational dressing is
that each measurement perturbs the local geometry by making
a small deficit angle $\alpha_k$ in the surface at the point
of measurement; there are no probes `outside' the universe
that can make such a measurement without perturbing the geometry.
It can and does happen that when gravity is switched on the
geometry is perturbed too violently by some operators
to have a good, local continuum limit.

In conclusion, we can take away the following main lessons
about Liouville theory coupled to conformal matter:

(1) KPZ scaling:
   $$\ZZ_g=C_g \mu^{(2-2g)Q/2\gamma}$$

(2) Operator scaling dimensions
$\OO_k^{\sst grav}=e^{\alpha_k\phi}\OO_k^{\sst matter}$
that shift the KPZ scaling relation by $\alpha_k$ in
correlation functions.  $-\hf(\alpha_k-\coeff Q2)^2+h_k=\frac{1-d}{24}$.

(3) A `phase transition' at $d=1$ where the gravitationally
dressed identity operator becomes tachyonic in the sense
described above.  There is no obvious reason why tree amplitudes might
not be analytically continued as in $d=26$ to yield a
sensible classical string S-matrix, but (as at $d=26$) loop amplitudes
are infinite because tachyons cause infrared divergences in
loop integrals even in Euclidean spacetime.  As emphasized
by Seiberg\seiberg,
this phase transition is not always
at $c=1$, but occurs whenever the spectrum has physical tachyons
$h_{min}<\frac{d-1}{24}$.

Properties (1) and (2) are special to gravitationally
dressed conformal matter; (3) is expected to be a generic feature
persisting even when the matter theory is massive.  One advantage
of the matrix model, to which we turn next, is the ability to
calculate the partition function and correlations even for
massive matter.

The KPZ scaling relations suggest that a large part of Liouville
dynamics is accounted for by the zero modes.  Also $c<1$ minimal
models coupled to gravity have the tachyon as the only
physical state (there are some additional physical states at
nonstandard values of the ghost number\ref\lz{B. Lian and
G. Zuckerman\journal Phys. Lett.&254B (91) 417.} whose meaning is less
clear); reparametrization invariance cancels the fluctuations
of the longitudinal modes, leaving only the center of mass motion
of the string when the spacetime is 1+1 dimensional.
Indeed, it has been shown that free string propagation
is accurately described by the quantum mechanics of the zero modes
\mss\foot{String interactions are not saturated by the
zero modes, which has been interpreted in \mss\ as due to the
violation of the single string physical state conditions by
contact interactions in the vertices.}
One can imagine solving the reparametrization invariance constraints
$T_{00}=T_{01}=0$ (or equivalently the BRST invariance constraints
in conformal gauge) to show that physical states contain
no excitations of the string's nonzero modes;
the remaining constraint is the Wheeler-deWitt equation
$T_{00}^{\sst zero\ mode}=0$
on the zero modes.  In a conformal matter theory coupled to gravity
this separates into the dynamics of the matter zero modes,
whose spectrum of scaling dimensions couples to the Liouville zero mode
equation\ct\seiberg\joe
$$\left[-(\ell\frac\partial{\partial\ell})^2+4\mu\ell^2+\nu^2\right]
	\Psi_\OO(\ell)=0  $$
with $\ell=e^{\gamma\phi/2}$ and $\nu=\frac2\gamma(\alpha-\frac Q2)$.
The appropriate solutions to this equation are the modified Bessel functions
\eqn\bessel{\Psi_\OO(\ell)=(\nu\sin\pi\nu)^{1/2}K_\nu(2\sqrt\mu\ell)  }
Physically, the function $K_\nu=\frac{\pi}{2\pi\sin\nu\pi}
[I_{-\nu}-I_\nu]$ is the linear combination of `incoming'
and `outgoing' waves $I_{\pm\nu}(2\sqrt\mu\ell)$ which is
exponentially damped in the infrared $\ell\rightarrow\infty$,
indicating total reflection.
Some of the strongest evidence for the viability of the conformal field
theory approach to 2d gravity is the appearance of these wavefunctions
within the matrix model.

\newsec{The matrix model}

The Hilbert space of Liouville is functionally infinite dimensional,
however the subspace of physical states is much smaller --
at most a countable number for $c\le1$.  Naively each gauge
invariance removes one canonical pair of variables, one by
a choice of gauge and another by the gauge constraint
(\ie\ that the generator of gauge transformations annihilate the
physical subspace -- the analogue of Gauss' law in QED).
In 2d gravity there are two reparametrization
degrees of freedom $\xi^a\rightarrow\tilde\xi^a(\xi)$.
The time components of the metric are the Lagrange multipliers
of the gauge constraints $T_{00}=T_{01}=0$; the canonical degrees of
freedom are the spatial metric $e^{\gamma\phi}$ and its
conjugate momentum $\pi_\phi$.  Thus we have the possibility
to gauge away up to one scalar matter fields' worth of local
degrees of freedom\foot{Note that in this counting the Ising
model has $d=\hf$ because $\pi_\psi\equiv\psi$ is `half'
a canonical pair.}.  As usual in
quantum gravity the difficulty in imposing the constraints
lies in the Hamiltonian constraint $T_{00}=0$, since this involves
the way that the 2d worldsheet is carved up into spacelike
hypersurfaces and therefore involves the (coordinate) time
development.  This makes it difficult to find nice global
gauge invariant states.  Nevertheless, the lesson to be drawn
is that it should pay to reformulate the path integral on the
space of physical configurations because it is expected to
be {\it much smaller} than the space of metrics plus matter
field configurations.  The simplest way to enumerate physical
configuarations is to discretize the 2d surfaces.  Consider
a 2d surface built by gluing together uniform squares
each with sides of length $\eps$ (or
triangles, pentagons, etc. -- it turns out not to matter
which\randsurf\ unless one artificially tunes couplings\ref\kaz{V.
Kazakov\journal Mod. Phys. Lett.&A4 (89) 2125.}).

\vbox to 2.5in{\vfil}
\centerline{{\it fig.4} Patch of discretized surface.}

This lattice spacing replaces the
covariant cutoff of Liouville theory.  The only local freedom
in pure gravity
resides in how many squares meet at each vertex, which determines
the local curvature discretized in units of $\pi/2$ (see fig.4).
Note that each surface is dual to a $\Phi^4$ graph (see fig.4),
each square being dual to a $\Phi^4$ vertex, each side of
the square being dual to a `propagator' of the $\Phi^4$
Feynman graph.
Thus counting all graphs with $A$ vertices counts all surfaces with
area $A\eps^2$; we can call this the partition function
for discrete 2d Euclidean quantum gravity.  The continuum limit consists
of taking $\eps\rightarrow0$, $A\rightarrow\infty$ such that
the physical area $A_{\sst phys}=A\eps^2$ is finite, assuming
such a limit exists.  In other words we concentrate on surfaces
with a very large number of triangles;
the statistics of these surfaces is governed by the large order
asymptotics of graphical perturbation theory.
We expect to be able to
generate any local curvature in the continuum by coarse-graining
over a large number of 3-, 4-, and $5^+$-coordinated vertices
on the dual tesselation, of positive,
zero and negative curvature, respectively.

In the pure gravity case,
the generating function for the $\Phi^4$ graphs dual to the
discretization is the integral
  \eqn\genfn{\int d\Phi\;e^{-\frac12\Phi^2+g\Phi^4}\ ,  }
\ie\ the coefficient of $g^A$ in the asymptotic expansion at small $g$
is the number of surfaces with area $A$ in lattice units,
since this asymptotic expansion {\it is} the enumeration of
$\Phi^4$ Feynman graphs.
The coupling $g$ is to be identified with the bare
2d cosmological constant $g=\exp[-\mu_{\sst bare}]$.  Note
that $g$ is positive in order that each surface is counted with
positive weight, so the generating function \genfn\ diverges
badly; we can interpret this through the asymptotic expansion
as the statement that the entropy of large surfaces diverges
uncontrollably.  To cut down on this entropy we can try to count
only surfaces of fixed genus, hoping that although the sum over
topology is infinite, the individual terms in the series might be
finite.  This turns out to be the case for $d\le1$ matter.

Matter can be incorporated by introducing a label set for the
dummy variable $\Phi$\ref\ising{V. Kazakov\journal Phys. Lett.&119A (86)
140; D. Boulatov and V. Kazakov\journal Phys. Lett.&186B (87) 379.};
then configurations are enumerated
not only by the connectivity of the graph but also the element
of the label set (which we think of as the value(s) of scalar
field(s) or of discrete spin variables) at each point on the graph.
For instance in the Ising model one considers two matrices $M$, $N$
with integrand
  $$\exp\bigl(-\tr[abM^2+\coeff ab N^2+cMN+gdM^4+\coeff gd N^4]\bigr)\ .$$
The couplings $a$, $b$ can be set to unity by a rescaling of
$M$, $N$ (however such redundant couplings can have physical
effects through contact terms in loop correlations\ref\mms{E. Martinec,
G. Moore, and N. Seiberg\journal Phys. Lett.&263B (91) 190.}).
That leaves $c$, related to the Ising temperature because it
controls the probability of transitions between up-spin ($M$)
and down-spin ($N$) subgraph domains in the diagrammatic expansion;
$d$ is related to the magnetic field since it preferentially weights
one of the two spins; and $g$ is our friend the surface cosmological
constant.
The trick in this and similar cases is to take the continuum limit cleverly
so that both graph connectivity and label fluctuations approach
criticality.
Fine-tuning more complicated potentials yields a tower of
critical points\ref\multi{T. Tada and M. Yamaguchi, UT-Komaba 90-17;
J. DeBoer, `Multimatrix models and the KP hierarchy',
Utrecht preprint THU 91/08.}.

To count the surfaces of fixed genus consider\ref\largen{E. Br\'ezin,
G. Parisi, C. Itzykson, and J.-B. Zuber\journal Comm. Math. Phys.&59 (79) 35;
C. Itzykson and J.-B. Zuber\journal J. Math Phys.&21 (80) 411;
D. Bessis, C. Itzykson, and J.-B. Zuber\journal Adv. Appl. Math.&1 (80) 109.}
the N$\times$N Hermitian
matrix field $\Phi_{\bar ab}$, and generating function
  $$\int d\Phi\;e^{-\tr[\frac12\Phi^2-(g/N)\Phi^4]}\ .  $$
The reason to make $\Phi$ Hermitian is that the Feynman graphs
carry an orientation and so the partition function
counts orientable surfaces only.  For the generalization to
unoriented surfaces, see \ref\unorient{G. Harris and E. Martinec\journal
Phys. Lett.&B245 (90) 384;
E. Brezin and H. Neuberger\journal Nucl. Phys.&B350 (91) 513.}.
Looking at simple graphs

\vbox to 2in{\vfil}

\noindent shows us that those which can be laid out smoothly on a surface
of genus $g$ come in the generating function with a factor
$N^{2-2g}$.  Each closed index loop traces over the
indices in that loop and gives a factor $N$;
the coupling constant of the generating
function is chosen to be $g/N$ so that adding a square without
changing the topology keeps the $N$ counting fixed (the extra $\frac1N$
cancels the additional index trace).
We have
  \eqn\matint{\ZZ_{\sst sft}=\int d\Phi\;e^{-N\tr[\frac12(\Phi/\sqrt N)^2
	-g(\Phi/\sqrt N)^4]}=\exp[-\sum_g N^{2-2g}C_g]\ .}
We have the right to call this generating function the
partition function of (discretized) string field theory in this
low-dimensional situation since it is indeed the object whose
asymptotic expansion in $N$ is the sum over surfaces
of some 2d field theory describing the string background.
The limit $N\rightarrow\infty$ picks out the sphere contribution
$C_{0}$, the classical limit of the associated string theory;
we can evaluate this from the integral \matint\ by saddle point
techniques even though the full integral doesn't make sense
(the saddle is not a global minimum) since the leading contribution
is just the value of the integrand at the saddle.  To find the
saddle, decompose the matrix in terms of `matrix polar coordinates' as
  $$\Phi=U\Lambda U^{-1}  $$
where $U$ is a unitary matrix and $\Lambda$ is the diagonal
matrix of eigenvalues of $\Phi$.  The partition function
becomes
  $$\ZZ_{sft}=\int U^{-1}dU\int d\Lambda
	\left|\frac{\p\Phi}{\p(U,\Lambda)}\right|\;\exp\left[-N
	\sum_{i=1}^N \bigl((\coeff{\lam_i}{\sqrt N})^2
		-g(\coeff{\lam_i}{\sqrt N})^4\bigr)\right]\ .$$
The Jacobian is easily evaluated by noting that a) it vanishes
whenever any two eigenvalues coincide; b) it is symmetric under
permutations of the eigenvalues; and c) it must scale
like $\lam^{N(N-1)}$; the unique function with these properties is
$\prod_{i<j}(\lam_i-\lam_j)^2$.  The justification for a) is analogous
to the vanishing of the Jacobian from Cartesian to polar coordinates
in ordinary multidimensional integrals:
the origin is a fixed point of the symmetry group of rotations.
The transformation is singular there so the measure must vanish.
Similarly, the coincidence of two eigenvalues is a fixed point
for an SU(2) subgroup of U($N$), so the Jacobian is singular.
The partition function becomes
  $$\ZZ_{sft}=\int\prod d\lam\;e^{-N\sum_i V(\lam_i/\sqrt N)+
	2\sum_{i<j}\log|(\lam_i-\lam_j)/\sqrt N|}\ .  $$
The mental picture to adopt is to consider the eigenvalues as
a set of particles lying in a metastable well, interacting through
a logarithmically repulsive `Coulomb' force.  As
$N\rightarrow\infty$ we can replace the set $\{\lam_i\}$
by an eigenvalue density (which we might regard as a
collective coordinate or string field\dj) and the saddle
point equation is
  $$\hf\lam-2g\lam^3=\int_{-a}^a\frac{\rho(\mu)d\mu}{\lam-\mu}\quad,
	\qquad\int_{-a}^a\rho(\mu)d\mu=1\ .  $$
Consideration of analytic properties provides the solution\largen
  $$\rho(\lam)=\coeff1\pi[\hf-ga^2-2g\lam^2]\sqrt{a^2-\lam^2}
	\qquad\quad\lam\le a=[\coeff{-1}{6g}[(1-48g)^{1/2}-1]^{1/2}]\ .$$
Near the endpoint $a$ of the distribution, $\rho(\lam)$
behaves as $\rho(\lam)\sim\sqrt{a^2-\lam^2}$; but when
$g\rightarrow g_c=-\frac1{48}$ the analytic behavior
changes to $\rho(\lam)\sim(a^2-\lam^2)^{3/2}$.
What is happening?  The edge of the eigenvalue density
becomes softer because the last eigenvalue approaches an instability
where the Coulomb repulsion of the rest of the eigenvalues
overcomes the external potential force and pushes it out
of the metastable well.  This instability reflects a phase
transition in the surface dynamics on the sphere: recall that
the entropy of surface configurations grows with the number of
plaquettes, but there is an energy cost $\exp[-\mu A]$.
At $g\sim g_c=e^{-\mu_c}$ from above these balance,
and $\ZZ$ is dominated by surfaces controllably large
compared to the cutoff $\eps$.  Inserting $\rho(\lam)$
into the saddle point action gives
  $$\eqalign{\coeff1{N^2}S_{\sst saddle}=&
	\int_{-a}^a\rho(\lam)V(\lam)
		-\int\int\rho(\mu)\rho(\lam)\log|\lam-\mu|\cr
	=&\coeff1{24}(a^2-1)(9-a^2)-\hf\log a^2\cr
	\sim& C_0(g-g_c)^{5/2} }$$
where $\eps^2\mu_{ren}=g-g_c$.
This saddle point action is the partition function of
2d gravity on the sphere, $S_{\sst saddle}=\ZZ^{2d}_{\sst sphere}
\sim C_0 N^2(\eps^2\mu_{ren})^{5/2}$.  Letting
  $$g_{str}=\frac1{N^2\eps^{5/2}}$$
gives $\ZZ_{\sst sphere}=C_0(\mu_{ren}^{5/2}/g_{str}^2)
\equiv C_0\mu^{5/2}$; in
general the full string partition function depends on the
couplings $\mu_{ren}$ and $g_{str}$ only through the
ratio $\mu$.  Note that to get a sensible continuum result we must let
$N$ scale with the surface cutoff $\delta\sim e^{\gamma\phi/2}$,
\ie\ $g_{str}$ is dynamical -- just what KPZ scaling predicts!
Plugging numbers into the Liouville formulae \soln,\kpzscal\
we find $Q=5\sqrt 3$, $\gamma=2/\sqrt 3$ for $d=0$, hence
Liouville theory predicts
  $$\ZZ_{\sst sphere}^{\sst Liouv}=C_0\mu^{\frac{Q}{2\gamma}\cdot 2}
	=C_0\mu^{5/2}$$
as observed!

Note that the important feature of the phase transition
used to take the continuum limit is the quadratic maximum
in the effective eigenvalue potential (linear vanishing
of the force on the last eigenvalue at $g=g_c$), so it shouldn't matter
whether we use $\Phi^4$ squares or $\Phi^3$ triangles in
the microscopic theory.  In other words, only the
rate of vanishing of $\rho$ near its endpoint is universal; the
details of the eigenvalue distribution far from this region as
well as the value of $g_c$ are irrelevant quantities.
By fine tuning the relative weight
of different polygonal simplices one can however reach other
sorts of critical points\kaz\ref\matt{M. Staudacher\journal Nucl. Phys.&B336
(90) 349.}.

The advance embodied in the discrete approach is the ease
with which one may calculate the coefficients $C_g$ in the
partition function\bkdsgm, as well as all correlation functions.
To date it has only been possible in the Liouville approach
to calculate (after much effort) correlation
functions on the sphere.  To go beyond the sphere one needs
to among other things incorporate corrections to $\rho(\lam)$
due to the discreteness of eigenvalues; $\rho$ is not a smooth
function at the $1/N$ level but rather a sum of delta functions.
In fact the whole methodology above is rather cumbersome, and it
proves simpler to reformulate the problem.  It is apparent that
the main difficulty is the Coulomb interaction among eigenvalues,
so we might try to find variables that
diagonalize this interaction as much as possible.
This happens\largen\
when one writes the square root of the
Jacobian as a Slater determinant
  $$\prod_{i<j}(\lam_i-\lam_j)=\det_{ij}[\psi_i(\lam_j)]
	\equiv|\Psi(\lam)\rangle \ , $$
where $\psi_i(\lam)=\lam^i-1+c_{i-2}\lam^{i-2}+\ldots+c_0$.
The $c_i$ are arbitrary since they correspond to $\psi_i\rightarrow
\psi_i+\psi_k$, $k<i$, which cancels from the determinant.
A convenient choice is to orthogonalize $\psi$ with respect to the
measure $e^{-V(\lam)}d\lam$
  $$\eqalign{\ZZ_{sft}=&\int\prod d\lam\bra{\Psi_\t(\vec\lam)}
	e^{-\sum_i V(\lam_i)}\ket {\Psi_\t(\vec\lam)}\equiv
	\bra{\Psi_\t(\vec\lam)}\left.\Psi_\t(\vec\lam)\right\rangle\cr
	V(\lam)=&\sum_k t_k\lam^k\cr}$$
Letting
\eqn\norm{
 h_n\delta_{nm}=\int d\lambda\ \psi_n(\lambda)
 e^{-V(\lambda)}\psi_m(\lambda)\ , }
we find
  $$\ZZ_{sft}=\prod_{i=1}^N h_i  $$
up to a $\t$-independent normalization.  The computation of the
partition function reduces to the determination of the $h_i(\vec t)$.
To this end, define
\eqn\innerprod{
  	h_nQ_{nm}=\int \psi_n(\lambda)
	e^{-V(\lambda)}\lambda \psi_m(\lambda) }
\eqn\deriv{
	h_nP_{nm}=\int \psi_n(\lambda) e^{-V(\lambda)}
	\frac d{d\lambda}\psi_m(\lambda)\ .  }
Considering different matrix elements yields
\eqn\jacobi{
\Q=\pmatrix{b_1&a_1&\  &\  &0 \cr
           1  &b_2&a_2&\  &\ \cr
           \  &1  &b_3&a_3&\ \cr
           0  &\  &\ddots&\ddots&\ddots}\ .
}
This also tells us (by acting $\lam$ to the left)
that $a_i=h_{i+1}/h_i$; a little thought shows $b_i=0$
if $V$ is an even function (since then we have the symmetry
$\psi_j(\lam)=(-)^j\psi(-\lam)$).  Integration by parts
for $\P$ gives
\eqn\pitoq{
        \P=V'(\Q)_+\ ,
}
where the $+$ subscript means taking only the upper triangular
part, setting the lower triangular and diagonal entries to
zero, since $d/d\lam$ must lower the order of $\psi$; similarly
a $-$ subscript refers to the lower triangular part.
Now
  \eqn\streqn{[\coeff d{d\lam},\lam]=1\Longrightarrow[\P,\Q]=\One }
is an equation for $a$, $b$ which solves our problem, \ie\
  $$\ZZ_{sft}=\prod_{i=1}^N h_i=h_0^N\exp[\sum_{i=1}^N
	(N-i)\log a_i] \ . $$
Afficionados of integrable systems will recognize that if $V=\hf\lam^2$
then $\P$, $\Q$ is the Lax pair of the Toda chain; actually
${\rm tr}\{Q^{k+1}\}$ is the $k^{\rm th}$ conserved charge of
the Toda problem and $(\Q^{k})_+$ is the $k^{\rm th}$ Toda
Hamiltonian.  In other words,
\eqn\floweqn{
\frac{\partial \Q}{\partial t_k}=[\Q^k_+,\Q]
}
are commuting flows.
One easily checks that these flows preserve the string equation \streqn\
Finally one can show that $\ZZ_{sft}$ is a particular
type of `tau-function' of the Toda system; this conveys little
information but sounds impressive, the content being primarily
in equations \streqn\ and \floweqn.  Commuting flows in this problem
should not surprise us; we must have $\frac{\p^2\ZZ}{\p t_i\p t_j}
=\frac{\p^2\ZZ}{\p t_j\p t_i}$, etc.,
so of course all the derivatives commute.  The nontrivial fact is
that we can represent $\ZZ$ as an inner product in a Hilbert space
\eqn\taufn{
        \ZZ_{sft}(\t)=
        \bra{\Psi_{\hat\t}} e^{-\sum (t_k-\hat t_k)\hat Q^k}
                \ket{\Psi_{\hat\t}}
}
where the derivatives with respect to the couplings act as operators.
The different flows in coupling space then form an integrable system
in this Hilbert space.

Again we need to take the continuum limit.  To this end it
is convenient to absorb a factor $e^{-V/2}$ in $\psi$,
normalize $\psi_k\rightarrow\frac1{\sqrt{h_k}}\psi_k$
and take $V$ even (although the latter is not essential).  Then
the Lax pair $\P$, $\Q$ are transformed into
  $$\qtilde=\Q_+-\Q_-\qquad\quad\ptilde=V'(\Q)_+-V'(\Q)_-\ . $$
The scaling limit consists of taking $a_i\rightarrow{\rm const.}+\eps^2
u(x)$, where $\frac iN=1-\eps^2x$.  The second order difference
operator $\qtilde$ becomes (in the sense of matrix elements)
a second order differential operator
$\p_x^2+u(x)$; $a_N\approx \frac{\ZZ_{N+1}\ZZ_{N-1}}{\ZZ_N^2}$
yields $u(x)=\p_x^2\FF$ where $\ZZ_{sft}=e^{-\FF}$.
Finally $\ptilde$ scales to some odd order differential operator
depending on the details of the potential (we made the rescaling in
the wavefunctions in order to make this manifest).
Generically $\ptilde\rightarrow\p_x$ which is boring, \ie
  $$[\ptilde,\qtilde]=\One=u'=\FF'''\Longrightarrow\FF=x^3  $$
which corresponds to $\rho\sim\sqrt{a^2-\lam^2}$.  Fine tuning
as before should yield $\rho\sim(a^2-\lam^2)^{3/2}$ corresponding
to $\ptilde\rightarrow \p^3+v_1(x)\p+v_0$.
Then the equation $[\ptilde,\qtilde]=1$
is the continuum statement
  $$u'''+uu'=1\ ,$$
the Painlev\'e I equation.  This equation embodies through its
asymptotic expansion in $x=\mu$ the entire perturbation series
for $\ZZ_{sft}$ about this background.  The leading solution
at large $x$ is $u^2=x$, which gives $\FF\sim x^{5/2}$
which is the famous KPZ scaling relation!  Plugging this solution
into the differential equation and iterating yields
  $$\FF=\sum_{g=0}^\infty C_g x^{\frac54(2-2g)}  $$
so we see $Q/\gamma=5/2$.
Note that derivative terms in the string equation come with factors
of the string loop coupling $g_{str}\sim 1/N$, which is why the genus
expansion expands in powers of derivatives of $u(x)$.
Douglas\douglas\ref\dfk{P. DiFrancesco and D. Kutasov\journal
Nucl. Phys.&342 (90) 589.} has generalized this construction for an arbitrary
pair of differential operators $\P$, $\Q$ of integer order $p$, $q$
satisfying \streqn, which produces BPZ minimal matter\bpz\
coupled to 2d gravity.  Changing the bare matrix potential scales
to continuum perturbations of $P$, $Q$ by lower or higher order
differential operators for relevant and irrelevant perturbations,
respectively\dfk\ref\effact{P. Ginsparg, M. Goulian, M.R. Plesser,
and J. Zinn-Justin\journal Nucl. Phys.&B342 (90) 539;
A. Jevicki and T. Yoneya\journal Mod. Phys. Lett.&A5 (90) 1615;
T. Yoneya, `Toward a canonical formalism of non-perturbative
2d gravity', U. Tokyo preprint UT-Komaba 91-8.}\foot{Of course any fixed
lattice operator is a sum over continuum operators with cutoff
dependent coefficients.}.
In the two (and more) matrix case, we can again eliminate the angle variables
in the partition function\largen.
To `diagonalize' the eigenvalue interaction, choose independently
the left and right wavefunctions $\overline{\bra{\Psi_\t}}$ and
$\ket{\Psi_\t}$\ref\mehta{S. Chadha, G. Mahoux, and M.L. Mehta\journal
J. Phys. A&14 (81) 579.}.
Defining matrices $M$, $P_M$ and $N$, $P_N$ analogous to
\innerprod\ and \deriv, we have the variational equations
$$\eqalign{P_M=&N+V_M'(M)\cr
	P_N=&M+V_N'(N)\cr}\ .$$
The number of nonzero diagonals of $M$, $N$ are determined by the
degrees of $V_N$, $V_M$ respectively. Tuning these potentials,
we can make $M$, $P_M$ scale to a pair of differential operators
$$\eqalign{Q=&\partial^q+u_{q-2}\partial^{q-2}+\ldots+u_0\cr
	   P=&\partial^p+v_{p-1}\partial^{p-1}+\ldots+v_0\cr}$$
of orders $(p,q)$.  The Heisenberg relation $[P_M,M]=1$
again determines the coefficients $u_n(x,\t)$, $v_m(x,\t)$
and hence the free energy $\partial_x^2\FF=u_{q-2}$\douglas\dfk.
These critical points have been identified with $(p,q)$ conformal
matter coupled to 2d gravity.  The flows in the couplings $\t$
are governed by the 2d Toda hierarchy on the lattice\discrete,
which scales to the $q$-reduced KP hierarchy in the
continuum\douglas\dfk\effact.
The coefficients $C_g\sim(2h)!$ for large $g$, reflecting the divergence
of the sum of the perturbation series\bkdsgm\bkdsgm.
It is very interesting that
these coefficients can be associated with an auxilliary, higher-action
saddle of the matrix integral\shenker\ corresponding to the eigenvalue
configuration where the last eigenvalue is moved to its
unstable equilibrium

\vbox to 2in{\vfil}

\noindent which might be interpreted as a `string instanton' mediating
some kind of vacuum decay, although the
precise meaning of this configuration remains unclear.

An interesting and calculable set of correlation functions in these
systems are correlations of the loop operator $W(\ell)$
which cuts a hole of boundary length $\ell$ in the surface.
In the matrix model this operator is $\frac1l\tr\Phi^l$,
which in graphs inserts a source of $l$ external lines
graphically dual to a loop or hole of boundary length $l\eps$.
For instance, fig.4 is a contribution to $l=24$.
In the continuum limit $\Q\rightarrow {\rm const.}+\eps(-\p^2+u)$,
$l\rightarrow\ell/\eps$
this becomes the heat operator
$e^{-\ell(-\p^2+U)}$
 $$\tr\Phi^l\sim({\rm const.}+\eps(-\p^2+u))^{\ell/\eps}
	\sim e^{-\ell(-\p^2+u)}\equiv W(\ell)$$
up to an unimportant (nonuniversal) normalization.
The calculation of loop correlation functions thus reduces to
a set of convolutions of heat kernels\bdss, \eg
\eqn\loops{\vev{W(\ell_1)W(\ell_2)}=\int_{-\mu}^{\infty}dx
	\int_{-\infty}^{-\mu}dy\bra x e^{-\ell_1Q}\ket y
	\bra y e^{-\ell_2Q}\ket x\ .}
For the general $(p,q)$ model we simply use the operator $Q$
determined by the string equations \streqn.
The loop correlations
serve as a kind of generating function of local operator
correlations, in the sense that the $\ell\rightarrow0$
limit of $W(\ell)$ can be recast as a sum (integral) over
the spectrum of local operators\mss.  Geometrically, the
zero mode $\ell=\oint e^{\gamma\phi/2}$ is dual (in the
sense of Fourier transform) to its conjugate
Liouville momentum $p=\oint \pi_\phi$ whose eigenvalue is the
deficit angle; hence it should be no surprise that a Green's
function at fixed small $\ell$ should be an integral transform
over local curvatures.
The small $\ell$ expansion of \loops\ is the asymptotic expansion
of the heat kernel, which is evaluated in terms of formal
fractional powers of $Q$\ref\gelfand{I. Gelfand and L. Dikii\journal
Funct. Anal. and Appl.&10 (76) 259; \andjournal Russ. Math. Surveys&30:5
(75) 77.}\bdss\mss.
For instance, \loops\ expands as
$$\eqalign{\vev{W(\ell)W(\ell')}\buildrel{\ell'\rightarrow0}\over{\sim}&
	\sum_{n=0}^\infty{\ell'}^{n/q}\vev{\OO_n^{KP}W(\ell)}\cr
	\vev{\OO_n^{KP}W(\ell)}=&\int^\mu dx\bigl[ Q_+^{n/q},
	\bra{x}e^{-\ell Q}\ket{x}\bigr]\ .\cr}$$
For the definition of fractional powers of differential operators
and their uses, see \eg\ \effact\dfk\ref\apfeldorf{K. Apfeldorf\journal
Nucl. Phys.&B360 (91) 480.}.
The basis of operators defined by isolating individual terms
on the RHS do not coincide with the dressed operators $\OO_\alpha$
of conformal field theory\mss.  Instead, one must take linear
combinations of $\OO_n^{KP}$ with coefficients defined by the
`incoming wave' $I_{n/q}(2\sqrt\mu\ell')$.
Then one finds the two-point function on the sphere
of a CFT scaling operator and a loop operator is
  \eqn\HH{\vev{\OO_\alpha W(\ell)}=(2\sqrt\mu)^\alpha
	K_\alpha(2\sqrt\mu\ell)\ .}
Thus one can justifiably claim that the matrix and Liouville
approaches coincide where calculations can be done in both.
Moreover, according to the ideas of Hartle and Hawking\hh,
one expects that the correlation function of a scaling operator
$\OO_k$ with a loop operator $W(\ell)$ is the corresponding
wavefunction of that operator -- the operator creates the appropriate
state in the Hilbert space of physical states, and $W(\ell)$ is
the probability of that state propagating on a Euclidean surface of
average curvature $-\mu$ to a boundary of length $\ell$.
This expectation is indeed borne out by the explicit calculations
\seiberg\mss\ outlined above.

The matrix model is an effective tool for the calculation of
correlations of integrated local scaling operators.  While this
is an important class of measurements it by no means
exhausts the list of interesting questions one can try to ask in 2d
gravity.  These integrated correlations are what might be called
extensive measurements, things like the area of the surface,
or the magnetization or
average energy density in the Ising model.
Truly local measurements, where one tries to `build a laboratory'
on a patch of the surface and make measurements in it,
have not been addressed yet because they are intrinsically more
complicated.  One must describe the laboratory in a coordinate invariant
way, \eg\ by prescribing the geodesic distances and relative
orientation of the objects in it.  Even the simplest such
`local' measurement, the Hausdorff dimension of the surface,
turns out to be very difficult to investigate\ref\david{F. David,
`What is the intrinsic geometry of 2d quantum gravity?',
Rutgers preprint RU-91-25.}.
It is of course possible that, like the scaling operators of
large positive Liouville momentum, such measurements in two dimensions
are plagued by strong fluctuations that render local questions
meaningless.  In such a situation the simple model does not
retain the flavor of its higher dimensional cousin.
Another interesting issue is whether we can define Minkowski
2d gravity independent of working in Minkowski string theory.
Usually we induce the continuation to Minkowski signature world
sheet through the analytic continuation of string amplitudes
from Euclidean spacetime and its associated $i\eps$
prescription.  Nevertheless it would be odd if the beautiful
physics of $1+1$-dimensional scattering theory
could not be put on a fluctuating geometry.  Can we not
discuss the S-matrix of the Ising model, or Potts or sine-gordon models,
in deSitter 2d gravity?  This also may hinge on how to
define localized asymptotic states in a fluctuating
background geometry.

\listrefs

\end